\documentclass{article}
\usepackage{spconf,amsmath,graphicx}
\usepackage{color,soul,cite}
\usepackage{MnSymbol}
\usepackage{wasysym}
\usepackage{multirow}
\usepackage{makecell}
\usepackage[rgb]{xcolor} 
\usepackage{float}
\usepackage{placeins}
\usepackage{psfrag}
\newcommand{\stz}{\rule{0mm}{2.3ex}}

\title{CONCATENATED IDENTICAL DNN (CI-DNN)\\ to REDUCE NOISE-TYPE DEPENDENCE IN DNN-Based SPEECH ENHANCEMENT}
\name{Ziyi Xu, Maximilian Strake, Tim Fingscheidt}
\address{Institute for Communications Technology, Technische Universit{\"a}t Braunschweig, Germany\\
$\left \{ \text{ziyi.xu, m.strake, t.fingscheidt} \right \}$@tu-bs.de}
\begin{document}
\ninept
\maketitle
\begin{abstract}
Estimating time-frequency domain masks for speech enhancement using deep learning approaches has recently become a popular field of research. In this paper, we propose a mask-based speech enhancement framework by using {\it concatenated identical} deep neural networks (CI-DNNs). The idea is that a single DNN is trained under multiple input and output signal-to-noise power ratio (SNR) conditions, using targets that provide a moderate SNR gain with respect to the input and therefore achieve a balance between speech component quality and noise suppression. We concatenate this single DNN several times without any retraining to provide enough noise attenuation. Simulation results show that our proposed CI-DNN outperforms enhancement methods using classical spectral weighting rules w.r.t.\ total speech quality and speech intelligibility. Moreover, our approach shows similar or even a little bit better performance with much fewer trainable parameters compared with a noisy-target single DNN approach of the same size. A comparison to the conventional clean-target single DNN approach shows that our proposed CI-DNN is better in speech component quality and much better in residual noise component quality. Most importantly, our new CI-DNN generalized best to an unseen noise type, if compared to the other tested deep learning approaches.
\end{abstract}
\vspace*{-4mm}
\section{Introduction}
\label{sec:intro}
\vspace*{-1mm}
Speech enhancement aims at improving the perceived quality and intelligibility of a speech signal degraded by additive noise. This task can be very challenging when only a single-channel mixture signal is available. The classical method to perform single-channel speech enhancement is to estimate the \textit{a priori} signal-to-noise ratio (SNR), which can subsequently be used by way of a spectral weighting rule \cite{Ephraim1984,Ephraim1985,Scalart1996,Lotter2005,Cohen2005a,Gerkmann2008b,Suhadi2011,Elshamy2015}. The decision-directed (DD) method proposed by Ephraim and Malah \cite{Ephraim1984}, or \cite{Cohen2005a,Gerkmann2008b}, are widespread \textit{a priori} SNR estimation approaches, that can be combined with spectral weighting rules such as the well known Wiener filter (WF) \cite{Scalart1996}, the MMSE log-spectral amplitude estimator (LSA) \cite{Ephraim1985}, and the super-Gaussian joint maximum \textit{a posteriori} estimator (SG) \cite{Lotter2005}. Nevertheless, using these classical approaches often still leads to poor performance in non-stationary noise environments \cite{malah1999tracking}.

Deep learning methods used in speech enhancement tasks have shown excellent results, even in non-stationary noise, and have become state of the art \cite{du2016regression,wang2014training,wang2015deep,erdogan2015phase,weninger2014discriminatively,williamson2016complex,xu2014experimental,wang2013towards,xu2015regression,elshamy2018dnn}. A regression-based speech enhancement method using deep neural networks is proposed in \cite{xu2014experimental}. Du et al. \cite{du2016regression} proposed a DNN architecture with dual outputs to estimate the speech features belonging to the target and interfering speakers from the input mixture signal, which achieves a better generalization to the unseen interfering speaker. Except of using DNNs for a direct regression task, a time-frequency domain mask can be estimated to perform speech enhancement making the ideal estimate independent of the absolute signal level \cite{wang2014training,wang2015deep,williamson2016complex}. A complex ratio mask that can enhance the amplitude spectrogram and estimate the right phase information is proposed in \cite{williamson2016complex}. Comparing with other mask estimation methods, using DNNs to directly predict the clean speech signal while estimating the mask representation implicitly is shown to outperform the direct estimation of masks for speech separation \cite{erdogan2015phase,weninger2014discriminatively}. For these deep learning based speech enhancement algorithms, a common problem is the degradation of performance in unseen noise conditions \cite{xu2014experimental,wang2013towards,xu2015regression}. One method to address this mismatched noise condition problem is to include many different noise types in the training data \cite{wang2013towards,xu2015regression}. A drawback of this method is that a very large training set is needed, e.g., Xu et al. \cite{xu2015regression} used 104 types of noise, and a total amount of 100 hours of training data for most experiments to improve generalization capabilities.

Another challenge in DNN-based speech enhancement is to find a good tradeoff between speech distortion and noise reduction, especially for low SNR conditions. Gao et al. \cite{gao2016snr} proposed to use progressive learning with SNR-based targets to address this problem. A novel progressive deep neural network (PDNN) with a parallel structure of neural networks with less and less noisy targets for each network and horizontal connections between the layers towards less-noisy trained DNNs is proposed in \cite{Shu2018iwaenc}. Training of this PDNN is done one-by-one while freezing the weights of the higher-noise target networks, until the last parallel network, which is trained with clean targets. The total spectral estimate output is the average of all these networks.

In this paper, however, we propose a {\it serial} concatenation of networks in so-called stages, with the specific property that the network in each stage is {\it identical}, hence the name {\it concatenated identical} DNNs (CI-DNNs). The idea is to train a basic DNN module which can yield some moderate enhancement of the input, in our case a $5\,$dB target signal-to-noise power ratio (SNR) improvement, implemented by using additive noise and a respectively configured target signal. This network is then, e.g., concatenated three times (i.e., 3 stages), in order to provide a sufficient amount of noise attenuation. An important aspect is that such a stage DNN must be trained for multiple input and (enhanced by $5\,$dB SNR improvement, respectively) output SNRs to operate well both in the first stage and in all subsequent stages. The idea and major advantage vs.\ \cite{gao2016snr,Shu2018iwaenc} is to save free (trainable) parameters compared to (deeper) DNNs with the same number of weights,  and thereby to provide better generalization properties particularly for unseen noise types, which so far is oftentimes reported to be a major issue in noise reduction by neural networks. 
\begin{figure*}[t!]
	\psfrag{A}[cc][cl]{$y(n)$}
	\psfrag{B}[cc][cc]{Window}
	\psfrag{C}[cc][cc]{$\&$ FFT}
	\psfrag{D}[cc][cr]{$Y_\ell(k)$}
	\psfrag{E}[cc][cc]{DNN enhancement}
	\psfrag{F}[cc][cc]{stage $r=1$}
	\psfrag{G}[cc][cc]{IFFT}
	\psfrag{H}[cc][cc]{$\&$ OLA}
	\psfrag{I}[cc][cc]{stage $r=2$}
	\psfrag{J}[cc][cc]{stage $r=R$}
	\psfrag{L}[cc][cr]{$\hat{S}_{1,\ell}\left (k \right )$}
	\psfrag{M}[cc][cc]{$\hat{S}_{2,\ell}\left (k \right )$}
	\psfrag{N}[cc][cc]{$\hat{S}_{R,\ell}\left (k \right )$}
	\psfrag{O}[cc][cr]{$\hat{s}(n)$}
	\centering
	\centerline{\includegraphics[width=17cm]{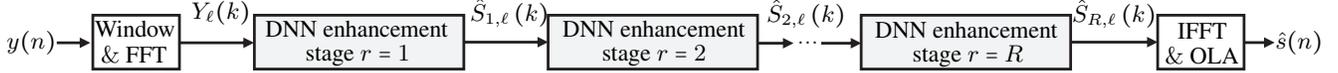}}
	\vspace*{-2mm}
	\caption{Diagram of the speech enhancement using CI-DNNs; for details of the DNN enhancement stage see Fig.\,2}
	\label{fig:res2}
	\vspace*{-1mm}
\end{figure*}
\begin{figure}[t!]
	\vspace*{-2mm}
	\psfrag{A}[cc][cr]{Input}
	\psfrag{B}[cc][cr]{$Y_\ell(k)$}
	\psfrag{C}[cc][cc]{$\left| Y_\ell(k) \right |$}
	\psfrag{D}[cc][cc]{$M_{\ell}(k)$}
	\psfrag{E}[cc][cc]{NORM}
	\psfrag{F}[cc][cc]{DNN}
	\psfrag{G}[cc][cc]{Output}
	\psfrag{H}[cc][cl]{$\hat{S}_{\ell}\left (k \right )$}
	\psfrag{I}[cc][cl]{(or training}
	\psfrag{J}[cc][cl]{target)}
	\psfrag{K}[cc][cr]{DNN enhancement stage}
	\centering
	\centerline{\includegraphics[width=8.5cm]{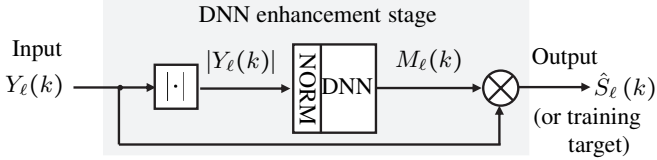}}
	\vspace*{-3mm}
	\caption{Basic DNN module for speech spectrum enhancement}
	\label{fig:res}
	\vspace*{-5mm}
\end{figure}

This paper is structured as follows: Section 2 describes our speech enhancement system and our novel CI-DNN architecture, along with training and testing aspects of the CI-DNN. The experimental setup as well as the results and discussion are presented in Section 3. We conclude the paper in Section 4.

\vspace*{-1mm}
\section{Concatenated Identical DNN}
\label{sec:format}
\vspace*{-2mm}
\subsection{Basic DNN Module and New CI-DNNs}
\label{ssec:sub1}
\vspace*{-1mm}
We assume the single-channel mixture $y(n)=s(n)+d(n)$ of the clean speech signal $s(n)$ and the added noise signal $d(n)$ with $n$ being the discrete-time sample index. Our speech enhancement system operates in the discrete Fourier transform (DFT) domain. Therefore, let $Y_\ell(k)$, $S_\ell(k)$, and $D_\ell(k)$ be the respective DFTs, and $\left |Y_\ell(k)\right |$, $\left |S_\ell(k)\right |$, and $\left |D_\ell(k)\right |$ be their DFT magnitudes, with frame index $\ell\in\mathcal{L}=\left\{1,2,\ldots,L\right\}$ and frequency bin index $k\in\mathcal{K}=\left \{ 0,1,\ldots,K\!-\!1 \right \}$ with DFT size $K$. In this paper, we only enhance the magnitude spectrogram of the noisy speech and use the unaltered noisy speech phase for reconstruction. Then, we can write our task as
\vspace*{-3mm}
\begin{equation} \label{clean_speech_est}
\hat{S}_\ell\left (k \right )=Y_\ell(k)\cdot M_\ell\left(k \right ),
\vspace*{-1mm}
\end{equation}
with $M_\ell\left(k \right )\in\left [ 0,1 \right ]$ and $\hat{S}_\ell\left (k \right )$ being the real-valued spectral mask and the estimated enhanced speech spectrum, respectively. As proposed in \cite{erdogan2015phase,weninger2014discriminatively}, we predict the unknown enhanced amplitude speech spectrum $\left |\hat{S}_\ell\left (k \right )\right |$, while estimating the spectral mask $M_\ell\left(k \right )$ implicitly as shown in Fig.\,\ref{fig:res}. The ``\,NORM\," operation in Fig.\,\ref{fig:res} represents a zero-mean and unit-variance normalization based on statistics collected on the training set.

Based on this single-stage basic DNN module, we build a speech enhancement system using the newly proposed serially concatenated identical DNN (CI-DNN) structure as shown in Fig.\,\ref{fig:res2}. Both topology and weights of each DNN enhancement stage are the same, in some more detail depicted as shown in Fig.\,\ref{fig:res}. The idea is to train a single basic DNN module which can offer some moderate enhancement of the input, in our case a $5\,$dB SNR improvement. Then we concatenate the same module several times in so-called {\it stages} with stage index $r\in\mathcal{R}=\left\{1,2,\ldots,R\right\}$ without any additional retraining. During inference, the DNN enhancement stage will enhance the input of stage $r$, and the output of stage $r$ serves as respective input for stage $r\!+\!1$. Hence, we divide our speech enhancement task into multiple sub-tasks, where the total number of stages can be decided by the total target SNR improvement. In this work, each stage is designed to improve the SNR by $5\,$dB, so the 2-stage and 3-stage CI-DNNs will ideally offer $10\,$dB and $15\,$dB SNR improvement, respectively. Another factor that can be influenced by the number of stages is the tradeoff between noise reduction and speech distortion. Being able to decide on the number of stages based on development set performance, without the need for retraining, makes our proposed CI-DNN very flexible to adapt for tasks with different requirements. The maximum number of stages we used for this work is $R=3$. The final stage output $\hat{S}_{R,\ell}\left (k \right )$ is transformed to the time domain by IFFT and overlap add (OLA) as shown in Fig.\,\ref{fig:res2}.

\vspace*{-3mm}
\subsection{New Approach Training}
\label{ssec:sub2}
\vspace*{-1mm}
The most important aspect to train a basic DNN module that can be concatenated as shown in Fig.\,\ref{fig:res2}, is to make sure the basic DNN is trained under multiple input and output SNR conditions to enable it to operate well both in the first stage and in all subsequent stages.

To train our basic module shown in Fig.\,\ref{fig:res}, we use the input noisy spectrogram $Y_\ell(k)$ in six SNR levels ranging from $-5\,$dB to $20\,$dB with a step size of $5\,$dB. The corresponding enhanced noisy targets $\hat{S}_\ell^{\text{target}}\left (k \right )$ have $5\,$dB higher SNR, which means the corresponding SNR levels range from $0\,$dB to $25\,$dB with the same $5\,$dB step size. The SNR level is measured according to ITU P.56 \cite{ITU56}. We define the loss function for each frame $\ell$ as
\vspace*{-1mm}
\begin{equation} \label{loss}
J_{\ell}=\frac{1}{K}\sum_{k\in\mathcal{K}}\left ( \left |\hat{S}_\ell\left (k \right )\right |-\left | \hat{S}_\ell^{\text{target}}\left (k \right ) \right | \right )^2,
\vspace*{-2.5mm}
\end{equation}
with $\left |\hat{S}_\ell\left (k \right )\right |$ being calculated using \eqref{clean_speech_est}. Given a DFT length of $K=256$, the input size of the basic DNN module is $5\times 129=645$, which includes 2 left and 2 right context frames. There are 5 hidden layers for each basic DNN module with a succeeding size of $1024-512-512-512-256$. All these hidden layers use leaky rectified linear units (ReLU) as activation function and a dropout rate of $p=0.2$. The size of the output layer is $\frac{K}{2}+1=129$, determined by our target dimensionality. We use a sigmoid activation function for the output layer to make sure the value of the mask $M_\ell\left(k \right )$ is between $0$ and $1$. All possible forward residual skip connections are added to the layers with matched dimensions, which results to 3 bypasses in total to ease the vanishing gradient problem during training \cite{veit2016residual}. Batch normalization is used for each layer except for the input layer, and we use a minibatch size of 128 for all trainings.

\vspace*{-3mm}
\subsection{New Approach Test}
\label{ssec:sub3}
\vspace*{-1mm}
As shown in Fig.\,\ref{fig:res2}, the input noisy speech spectrum $Y_\ell(k)$ will be enhanced progressively as
\vspace*{-2mm}
\begin{equation} \label{progre_enhance}
\hat{S}_{R,\ell}\left (k \right )=Y_\ell(k)\cdot \prod_{r=1}^{R}M_{r,\ell}(k),
\vspace*{-1.5mm}
\end{equation}
with $M_{r,\ell}(k)$ being the estimated mask in stage $r$. Since the identical basic DNN module uses the same number of context frames in each stage, an additional amount of context is needed the more stages are employed. As an example, the 2-stage CI-DNN needs 9 frames at the input of the first stage, which includes 4 left and 4 right context frames to produce one output frame at the final stage. This number will increase to 13 frames including 6 left and 6 right context frames for a 3-stage CI-DNN.

\vspace*{-1mm}
\section{Experimental Validation}
\label{sec:majhead}
\vspace*{-2mm}
\subsection{Database and Measures}
\label{ssec:sub3_1}
\vspace*{-1mm}
The clean speech data for our training and test is taken from the Grid Corpus \cite{grid_corpus}. To make our CI-DNNs {\it speaker-independent}, we randomly select 16 speakers, containing 8 male and 8 female speakers, and use 160 sentences per speaker for the basic DNN module training. For evaluation, four different speakers are chosen, two male and two female, with 10 sentences each.

Three types of superimposed noise are used to construct the training data: Pedestrian noise (PED), caf\'e noise (CAFE), and street noise (STR) which are obtained from the CHiME-3 dataset \cite{chime3}. We train the basic DNN module under multiple input and output SNR conditions containing 6 different SNR levels. From the overall training material, $20\%$ of the data is used for validation and $80\%$ is used for actual training. All the speech and noise signals have a sampling rate of $16\, \text{kHz}$ and are transferred to the DFT domain with $K=256$ using a periodic Hann window with $50\%$ overlap.

The test data is constructed using PED and CAFE noise, however, extracted from different files. Speech material is from unseen speakers. To additionally perform a {\it noise-type independent} test, we also create test data using bus noise (BUS), taken also from the CHiME-3 data, with this noise type not being seen during training. All the test data sets contain SNR levels from $-5\, \text{dB}$ to $20\, \text{dB}$ with a step size of $5\,\text{dB}$. The evaluation is based on both the {\it filtered} clean speech component $\tilde{s}(n)$, the {\it filtered} noise component $\tilde{d}(n)$, and also the enhanced speech signal $\hat{s}(n)$. Using \eqref{progre_enhance}, $\tilde{S}\left ( \ell,k \right )$ and $\tilde{D}\left ( \ell,k \right )$ are obtained, replacing $Y_\ell(k)$ by $S_\ell(k)$ and $D_\ell(k)$, respectively.

In this paper, we use the following measures \cite{samy_SNR}:\\
1) SNR improvement: $\Delta\text{SNR}=\text{SNR}_{\text{out}}-\text{SNR}_{\text{in}}$, measured in $\text{dB}$\\
2) Speech quality (PESQ MOS-LQO) is measured using $s(n)$ as reference signal and either the {\it filtered} clean speech component $\tilde{s}(n)$ or the enhanced speech $\hat{s}(n)$ as test signal according to \cite{ITU862,ITU1110}, being referred to as PESQ$(\tilde{s})$ and PESQ$(\hat{s})$, respectively.\\
3) Segmental speech-to-speech-distortion ratio:
\vspace*{-2mm}
\[\text{SSDR}=\frac{1}{\left |\mathcal{L}_1 \right |}\sum _{\ell\in\mathcal{L}_1}\text{SSDR}(\ell)\qquad[\text{dB}]\vspace*{-2mm} \]
\vspace*{+0.5mm}
with $\mathcal{L}_1\subset\mathcal{L}$, denoting the set of speech-active frames \cite{samy_SNR}, and using $\text{SSDR}(\ell)=\max\left \{ \min\left \{ \text{SSDR}'(\ell),30\,\text{dB} \right \}, -10\,\text{dB} \right \}, \text{with}$
$\text{SSDR}'\!(\ell)\!=\!10\log_{10}\!\big(\!\big( \sum\limits_{n\in\mathcal{N}_\ell}\! s^2(n)\big)\!/\! ( \sum\limits_{n\in\mathcal{N}_\ell}\! \left [\! \tilde{s}(n+\Delta)\!-\!s(n)\right ]^2  \big)\!\big),$\vspace*{1mm}
with $\mathcal{N}_\ell$ denoting the sample indices $n$ in frame $\ell$, and $\Delta$ being used to perform time alignment for the filtered signal $\tilde{s}(n)$.\\
4) The weighted log-average kurtosis ratio (WLAKR) measures the noise distortion (especially for musical tones) using $d(n)$ as reference signal and the {\it filtered} noise component $\tilde{d}(n)$ as test signal according to ITU P.1130 \cite{ITU1130,yuDSP2011a}. A WLAKR score that is closer to zero indicates less noise distortion, whereas being far away (+ or -) from zero indicates strong noise distortion \cite{yuDSP2011a}. In our analysis we will show averaged {\it absolute} WLAKR values.\\
5) Short-time objective intelligibility (STOI) measures the intelligibility of the enhanced speech as proposed in \cite{taal2010short}.

We group these measurements to noise component measures ($\Delta$SNR and WLAKR), speech component ones (SSDR and PESQ$(\tilde{s})$), and total performance measures (PESQ$(\hat{s})$ and STOI).

\vspace*{-3mm}
\subsection{Baseline Methods}
\label{ssec:sub3_2}
\vspace*{-1mm}
The baseline methods include the classical LSA, SG, and WF spectral weighting rules combined with the DD approach for \textit{a priori} SNR estimation and  minimum statistics (MS) \cite{Martin2001} for noise power estimation as mentioned before. To compare with the conventional mask estimation methods, we also train a so-called single DNN for speech enhancement similar to Fig.\,\ref{fig:res}, but potentially deeper. We construct three single DNNs named 1 stage (same as CI-DNN, 1.23M weights), ``\,2 stage\," (2.43M weights), and ``\,3 stage\," (4.55M weights). These three single DNNs have a similar number of weights compared to the 1-stage, 2-stage, and 3-stage CI-DNNs, respectively, but all the weights in these baseline single DNNs are free weights that are trainable.

As mentioned before, the number of input frames required to obtain one output frame is different for CI-DNNs with different numbers of stages. To allow a fair comparison, the input size of the ``\,2 stage\," and ``\,3 stage\," single large DNNs are $9\times 129=1161$ with 4 left and 4 right context frames, and $13\times 129=1677$ with 6 left and 6 right context frames, respectively.

There are 6 hidden layers for the ``\,2 stage\," single large DNN with sizes of $1400-800-512-512-512-256$. The ``\,3 stage\," single large DNN contains 7 hidden layers with sizes of $1800-750-512-512-512-512-256$. Both networks have an output layer size of 129 using sigmoid activation functions. Except for the output layer, all layers use the same leaky ReLU activation functions. The same dropout rate $p=0.2$ is used in all hidden layers and batch normalization is used except for the input layer. All possible bypasses are employed, which results in a total of 3 and 6 bypasses for ``\,2 stage\," and ``\,3 stage\,", respectively. The 1 stage network has the same structure as our basic DNN module with identical weights.

We train these three baseline single DNNs using either clean speech or noisy speech as targets, separately. For the noisy target training, the target noisy speech provides $5\,$dB, $10\,$dB, and $15\,$dB SNR improvement for the 1 stage, ``\,2 stage\,", and ``\,3 stage\," single DNNs, respectively. 

\vspace*{-3mm}
\subsection{Experimental Results and Discussion}
\label{ssec:sub3_3}
\vspace*{-1mm}
We report on PED noise, CAFE noise, and on unseen BUS noise separately, both for all baselines and our proposed CI-DNNs. The measures are averaged over all speakers and all SNR levels, and are shown in Tables\ 1, 2, and 3. Additionally, we report the performance at $\text{SNR}=-5\,$dB, where results are shown in Tables\ 4, 5, and 6. In each column, the \textbf{two best results} are {\bf \colorbox{black!15}{greyshaded}}.

First, we look at PED noise and CAFE noise which are the {\it seen} noise types (not seen noises!) in CI-DNN and single DNN training.

The classical spectral weighting rule baselines LSA, SG, WF are strong in speech component quality, particularly at low SNR, which is reflected in very good PESQ$(\tilde{s})$ performance as shown in Table\ 4 and 5. However, they are all very bad in terms of residual noise quality (the by far highest WLAKR values for this measurement in all tables), and in terms of speech intelligibility, which shows an almost $0.1$ points lower STOI score compared to the other methods.

Inspecting the single DNN trained with clean speech target, we find that this method always shows strong performance in $\Delta$SNR and in total quality measures (see Tables\,1, 2, 4, and 5), but low fidelity of the speech component measures (SSDR and PESQ$(\tilde{s})$), interestingly. Although PESQ$(\hat{s})$ accepts this due to the high noise suppression, the clean target training seems to provide unbalanced results. Moreover, the residual noise quality, particularly at low SNRs, suffers audibly, which is also reflected by more than ten times higher WLAKR scores compared to other DNN-based or CI-DNN-based methods.

Now, we discuss the single DNN trained with noisy speech targets and our proposed CI-DNN (comparisons between DNNs and CI-DNNs with a similar number of weights). The single DNN trained with noisy target shows very good STOI, speech and noise component quality --- as is the case with our new CI-DNN. Regarding the speech component measures, the noisy-target single DNN is a bit better in SSDR, while CI-DNN is a bit better in PESQ$(\tilde{s})$. The noisy-target single DNN provides a bit more $\Delta$SNR in very noisy condition and on average, but not consistently so. Finally, concerning total PESQ$(\hat{s})$, CI-DNN is always (at low SNR, and average over SNR conditions) {\it slightly ahead} of the DNN with noisy-target training (for two or more stages, of course). Our proposed CI-DNNs only have $1/2$ or $1/3$ of the trainable weights compared to the noisy-target single DNN to achieve this performance. In summary, {\it for noise types that have been seen in training, the CI-DNN is overall slightly ahead of any here investigated baseline single DNN in terms of total speech quality.}

Secondly, we look at the measurement results for the {\it unseen} noise type BUS. As expected, the classical spectral weighting rule baselines perform similarly bad in BUS noise concerning background noise quality (WLAKR scores) and total speech intelligibility (STOI). For the neural network approaches, the BUS noise type has not been seen in training. They all perform nicely and equally well in STOI as shown in Tables\,3 and 6. The relative performance of the DNN trained with noisy targets vs.\ the DNN trained with clean targets is the same, whether we test with seen or unseen noise types, again disqualifying the clean target training due to its bad residual noise quality performance. 
\begin{table}[t!]
	\Large
	\centering
	\resizebox{1\linewidth}{!}{
		\begin{tabular}{|c|c||c|c||c|c||c|c|}
			\hline
			\multicolumn{2}{|c||}{\multirow{2}{*}{Method}}                                                                           & \multicolumn{2}{c||}{Noise Component} & \multicolumn{2}{c||}{Speech Component} & \multicolumn{2}{c|}{Total}        \\ \cline{3-8} 
			\multicolumn{2}{|c||}{}                                                                                                  & \stz{$\Delta$SNR}       & WLAKR            & SSDR              & PESQ$(\tilde{s})$  & PESQ$(\hat{s})$  & STOI            \\ \Xhline{3\arrayrulewidth}
			\multicolumn{2}{|c||}{\stz{LSA}}                                                                                               & 3.08              & 0.66             & \colorbox{black!15}{$\mathbf{15.05}$}  & 3.40              & 2.09            & 0.62            \\ \hline
			\multicolumn{2}{|c||}{\stz{SG}}                                                                                                & 2.73              & 0.70             & \colorbox{black!15}{$\mathbf{15.33}$}  & 3.37              & 2.06            & 0.62            \\ \hline
			\multicolumn{2}{|c||}{\stz{WF}}                                                                                                & 3.85              & 0.75             & 13.52             & 3.45              & 2.09            & 0.61            \\ \hline
			\multirow{3}{*}{\begin{tabular}[c]{@{}c@{}}Single DNN\\ clean target\end{tabular}}                                & \stz{1 stage}    & 6.67              & 0.11             & 13.24             & 3.20              & 2.45            & \colorbox{black!15}{$\mathbf{0.72}$} \\ \cline{2-8} 
			& \stz{``2 stage"} & \colorbox{black!15}{$\mathbf{6.71}$}   & 0.28             & 13.33             & 3.20              & \colorbox{black!15}{$\mathbf{2.55}$} & \colorbox{black!15}{$\mathbf{0.72}$} \\ \cline{2-8} 
			& \stz{``3 stage"} & \colorbox{black!15}{$\mathbf{6.78}$}   & 0.20             & 13.46             & 3.21              & \colorbox{black!15}{$\mathbf{2.51}$} & \colorbox{black!15}{$\mathbf{0.72}$} \\ \hline
			\multirow{3}{*}{\begin{tabular}[c]{@{}c@{}}Single DNN\\ noisy target\end{tabular}}                                & \stz{1 stage}    & 2.99              & \colorbox{black!15}{$\mathbf{0.02}$}  & 14.24             & \colorbox{black!15}{$\mathbf{3.52}$}   & 2.11            & 0.70            \\ \cline{2-8} 
			& \stz{``2 stage"} & 5.15              & \colorbox{black!15}{$\mathbf{0.02}$}  & 13.87             & \colorbox{black!15}{$\mathbf{3.48}$}   & 2.25            & \colorbox{black!15}{$\mathbf{0.71}$} \\ \cline{2-8} 
			& \stz{``3 stage"} & 6.40              & 0.05             & 13.73             & 3.40              & 2.37            & \colorbox{black!15}{$\mathbf{0.72}$} \\ \hline
			\multirow{3}{*}{\begin{tabular}[c]{@{}c@{}}New CI-DNN\\  $\Delta$SNR target: +5dB\\ for each stage\end{tabular}} & \stz{1 stage}    & 2.99              & \colorbox{black!15}{$\mathbf{0.02}$}  & 14.24             & \colorbox{black!15}{$\mathbf{3.52}$}   & 2.11            & 0.70            \\ \cline{2-8} 
			& \stz{2 stage}    & 5.07              & \colorbox{black!15}{$\mathbf{0.02}$}  & 13.58             & \colorbox{black!15}{$\mathbf{3.52}$}   & 2.28            & \colorbox{black!15}{$\mathbf{0.71}$} \\ \cline{2-8} 
			& \stz{3 stage}    & 6.03              & \colorbox{black!15}{$\mathbf{0.03}$}  & 12.74             & 3.47              & 2.43            & \colorbox{black!15}{$\mathbf{0.71}$} \\ \hline
		\end{tabular}
	}
	\vspace*{-4mm}
	\caption{Performance for PED noise, {\bf all SNRs} averaged; $\Delta$SNR and SSDR are measured in dB. {\bf Two best are \colorbox{black!15}{greyshaded}}.}
	\label{PED-noise}
	\vspace*{-3mm}
\end{table}
\begin{table}[t!]
	\Large
	\centering
	\resizebox{1\linewidth}{!}{
		\begin{tabular}{|c|c||c|c||c|c||c|c|}
			\hline
			\multicolumn{2}{|c||}{\multirow{2}{*}{Method}}                                                                           & \multicolumn{2}{c||}{Noise Component} & \multicolumn{2}{c||}{Speech Component} & \multicolumn{2}{c|}{Total}        \\ \cline{3-8} 
			\multicolumn{2}{|c||}{}                                                                                                  & \stz{$\Delta$SNR}       & WLAKR            & SSDR              & PESQ$(\tilde{s})$  & PESQ$(\hat{s})$  & STOI            \\ \Xhline{3\arrayrulewidth}
			\multicolumn{2}{|c||}{\stz{LSA}}                                                                                               & 3.48              & 0.61             & \colorbox{black!15}{$\mathbf{14.43}$}  & 3.40              & 2.17            & 0.63            \\ \hline
			\multicolumn{2}{|c||}{\stz{SG}}                                                                                                & 3.29              & 0.65             & \colorbox{black!15}{$\mathbf{14.65}$}  & 3.38              & 2.14            & 0.62            \\ \hline
			\multicolumn{2}{|c||}{\stz{WF}}                                                                                                & 4.30              & 0.68             & 12.90             & \colorbox{black!15}{$\mathbf{3.48}$}   & 2.18            & 0.62            \\ \hline
			\multirow{3}{*}{\begin{tabular}[c]{@{}c@{}}Single DNN\\ clean target\end{tabular}}                                & \stz{1 stage}    & 5.73              & 0.11             & 12.87             & 3.22              & 2.52            & \colorbox{black!15}{$\mathbf{0.71}$} \\ \cline{2-8} 
			& \stz{``2 stage"} & \colorbox{black!15}{$\mathbf{5.99}$}   & 0.13             & 12.94             & 3.19              & \colorbox{black!15}{$\mathbf{2.58}$} & \colorbox{black!15}{$\mathbf{0.71}$} \\ \cline{2-8} 
			& \stz{``3 stage"} & \colorbox{black!15}{$\mathbf{6.01}$}   & 0.11             & 13.02             & 3.19              & \colorbox{black!15}{$\mathbf{2.55}$} & \colorbox{black!15}{$\mathbf{0.72}$} \\ \hline
			\multirow{3}{*}{\begin{tabular}[c]{@{}c@{}}Single DNN\\ noisy target\end{tabular}}                                & \stz{1 stage}    & 2.72              & \colorbox{black!15}{$\mathbf{0.02}$}  & 13.71             & \colorbox{black!15}{$\mathbf{3.48}$}   & 2.19            & \colorbox{black!15}{$\mathbf{0.71}$} \\ \cline{2-8} 
			& \stz{``2 stage"} & 4.62              & \colorbox{black!15}{$\mathbf{0.02}$}  & 13.39             & \colorbox{black!15}{$\mathbf{3.45}$}   & 2.33            & \colorbox{black!15}{$\mathbf{0.71}$} \\ \cline{2-8} 
			& \stz{``3 stage"} & 5.78              & 0.04             & 13.29             & 3.37              & 2.44            & \colorbox{black!15}{$\mathbf{0.72}$} \\ \hline
			\multirow{3}{*}{\begin{tabular}[c]{@{}c@{}}New CI-DNN\\  $\Delta$SNR target: +5dB\\ for each stage\end{tabular}} & \stz{1 stage}    & 2.72              & \colorbox{black!15}{$\mathbf{0.02}$}  & 13.71             & \colorbox{black!15}{$\mathbf{3.48}$}   & 2.19            & \colorbox{black!15}{$\mathbf{0.71}$} \\ \cline{2-8} 
			& \stz{2 stage}    & 4.54              & \colorbox{black!15}{$\mathbf{0.03}$}   & 13.04             & \colorbox{black!15}{$\mathbf{3.48}$}   & 2.36            & \colorbox{black!15}{$\mathbf{0.71}$} \\ \cline{2-8} 
			& \stz{3 stage}    & 5.97              & \colorbox{black!15}{$\mathbf{0.03}$}   & 12.17             & 3.41              & 2.50            & \colorbox{black!15}{$\mathbf{0.71}$} \\ \hline
		\end{tabular}
	}
	\vspace*{-4mm}
	\caption{Performance for CAFE noise, {\bf all SNRs} averaged; $\Delta$SNR and SSDR are measured in dB. {\bf Two best are \colorbox{black!15}{greyshaded}}.}
	\label{CAFE-noise}
	\vspace*{-3mm}
\end{table}
\begin{table}[t!]
	\Large
	\centering
	\resizebox{1\linewidth}{!}{
		\begin{tabular}{|c|c||c|c||c|c||c|c|}
			\hline
			\multicolumn{2}{|c||}{\multirow{2}{*}{Method}}                                                                           & \multicolumn{2}{c||}{Noise Component} & \multicolumn{2}{c||}{Speech Component} & \multicolumn{2}{c|}{Total}        \\ \cline{3-8} 
			\multicolumn{2}{|c||}{}                                                                                                   & \stz{$\Delta$SNR}     & WLAKR             & SSDR              & PESQ$(\tilde{s})$  & PESQ$(\hat{s})$  & STOI            \\ \Xhline{3\arrayrulewidth}
			\multicolumn{2}{|c||}{\stz{LSA}}                                                                                               & 4.68             & 0.56              & \colorbox{black!15}{$\mathbf{16.08}$}  & 3.57              & \colorbox{black!15}{$\mathbf{2.71}$} & 0.72            \\ \hline
			\multicolumn{2}{|c||}{\stz{SG}}                                                                                                & 3.57             & 0.63              & \colorbox{black!15}{$\mathbf{16.21}$}  & \colorbox{black!15}{$\mathbf{3.62}$}   & \colorbox{black!15}{$\mathbf{2.69}$} & 0.72            \\ \hline
			\multicolumn{2}{|c||}{\stz{WF}}                                                                                                & \colorbox{black!15}{$\mathbf{6.24}$}  & 0.68              & 14.62             & \colorbox{black!15}{$\mathbf{3.76}$}   & \colorbox{black!15}{$\mathbf{2.71}$} & 0.71            \\ \hline
			\multirow{3}{*}{\begin{tabular}[c]{@{}c@{}}Single DNN\\ clean target\end{tabular}}                                & \stz{1 stage}    & 5.01             & 0.06             & 15.42             & 3.36              & 2.58            & \colorbox{black!15}{$\mathbf{0.77}$} \\ \cline{2-8} 
			& \stz{``2 stage"} & 3.95             & 0.17             & 15.45             & 3.33              & 2.57            & \colorbox{black!15}{$\mathbf{0.77}$} \\ \cline{2-8} 
			& \stz{``3 stage"} & 3.78             & 0.13             & 15.54             & 3.34              & 2.55            & \colorbox{black!15}{$\mathbf{0.77}$} \\ \hline
			\multirow{3}{*}{\begin{tabular}[c]{@{}c@{}}Single DNN\\ noisy target\end{tabular}}                                & \stz{1 stage}    & 1.90             & \colorbox{black!15}{$\mathbf{0.03}$}  & 15.96             & 3.47              & 2.35            & \colorbox{black!15}{$\mathbf{0.77}$} \\ \cline{2-8} 
			& \stz{``2 stage"} & 2.90             & 0.05             & 15.78             & 3.46              & 2.48            & \colorbox{black!15}{$\mathbf{0.77}$} \\ \cline{2-8} 
			& \stz{``3 stage"} & 3.67             & 0.08             & 15.71             & 3.41              & 2.54            & \colorbox{black!15}{$\mathbf{0.77}$} \\ \hline
			\multirow{3}{*}{\begin{tabular}[c]{@{}c@{}}New CI-DNN\\  $\Delta$SNR target: +5dB\\ for each stage\end{tabular}} & \stz{1 stage}    & 1.90             & \colorbox{black!15}{$\mathbf{0.03}$}  & 15.96             & 3.47              & 2.35            & \colorbox{black!15}{$\mathbf{0.77}$} \\ \cline{2-8} 
			& \stz{2 stage}    & 4.02             & \colorbox{black!15}{$\mathbf{0.04}$}  & 15.18             & 3.49              & 2.56            & \colorbox{black!15}{$\mathbf{0.77}$} \\ \cline{2-8} 
			& \stz{3 stage}    & \colorbox{black!15}{$\mathbf{5.86}$}  & 0.05             & 14.14             & 3.44              & \colorbox{black!15}{$\mathbf{2.71}$} & \colorbox{black!15}{$\mathbf{0.76}$} \\ \hline
		\end{tabular}
	}
	\vspace*{-4mm}
	\caption{Performance for {\bf unseen} BUS noise, {\bf all SNRs} averaged; $\Delta$SNR and SSDR are measured in dB. {\bf Two best are \colorbox{black!15}{greyshaded}}.}
	\label{BUS-noise}
	\vspace*{-6mm}
\end{table}

Comparing the noisy-target single DNN to the new CI-DNN in the unseen BUS noise type, however, we make surprising observations: The speech component quality is roughly comparable as before (SSDR better for DNN with noisy target, PESQ$(\tilde{s})$ better with CI-DNN), so is also the noise component quality (WLAKR). {\it However, the 3-stage CI-DNN clearly excels the respective single DNN in $\Delta$SNR. In the 3-stage case, the total PESQ$(\hat{s})$ of the CI-DNN is on average over all SNRs by 0.17 points better than that for the single DNN.}

\vspace*{-7mm}
\section{Conclusions}
\label{sec:con}
\vspace*{-2mm}
\begin{table}[t!]
	\Large
	\centering
	\resizebox{1\linewidth}{!}{
		\begin{tabular}{|c|c||c|c||c|c||c|c|}
			\hline
			\multicolumn{2}{|c||}{\multirow{2}{*}{Method}}                                                                           & \multicolumn{2}{c||}{Noise Component} & \multicolumn{2}{c||}{Speech Component} & \multicolumn{2}{c|}{Total}        \\ \cline{3-8} 
			\multicolumn{2}{|c||}{}                                                                                                    & \stz{$\Delta$SNR}     & WLAKR            & SSDR              & PESQ$(\tilde{s})$  & PESQ$(\hat{s})$  & STOI            \\ \Xhline{3\arrayrulewidth}
			\multicolumn{2}{|c||}{\stz{LSA}}                                                                                               & 2.22              & 0.66             & \colorbox{black!15}{$\mathbf{5.86}$}   & 2.54              & 1.34            & 0.41            \\ \hline
			\multicolumn{2}{|c||}{\stz{SG}}                                                                                                & 2.02              & 0.66             & \colorbox{black!15}{$\mathbf{5.47}$}   & \colorbox{black!15}{$\mathbf{2.72}$}   & 1.31            & 0.40            \\ \hline
			\multicolumn{2}{|c||}{\stz{WF}}                                                                                                & 2.41              & 0.75             & 4.42              & \colorbox{black!15}{$\mathbf{2.87}$}   & 1.33            & 0.41            \\ \hline
			\multirow{3}{*}{\begin{tabular}[c]{@{}c@{}}Single DNN\\ clean target\end{tabular}}                                & \stz{1 stage}    & 6.16              & 0.15             & 3.96              & 2.21              & 1.57            & \colorbox{black!15}{$\mathbf{0.54}$}  \\ \cline{2-8} 
			& \stz{``2 stage"} & \colorbox{black!15}{$\mathbf{7.00}$}   & 0.27             & 4.02              & 2.20              & \colorbox{black!15}{$\mathbf{1.56}$} & \colorbox{black!15}{$\mathbf{0.54}$} \\ \cline{2-8} 
			& \stz{``3 stage"} & \colorbox{black!15}{$\mathbf{7.21}$}   & 0.30             & 4.06              & 2.18              & \colorbox{black!15}{$\mathbf{1.57}$} & \colorbox{black!15}{$\mathbf{0.55}$} \\ \hline
			\multirow{3}{*}{\begin{tabular}[c]{@{}c@{}}Single DNN\\ noisy target\end{tabular}}                                & \stz{1 stage}    & 2.18              & \colorbox{black!15}{$\mathbf{0.02}$}  & 4.88              & 2.51              & 1.41            & \colorbox{black!15}{$\mathbf{0.54}$} \\ \cline{2-8} 
			& \stz{``2 stage"} & 4.57              & \colorbox{black!15}{$\mathbf{0.02}$}  & 4.53              & 2.46              & 1.45            & \colorbox{black!15}{$\mathbf{0.55}$} \\ \cline{2-8} 
			& \stz{``3 stage"} & 6.26              & \colorbox{black!15}{$\mathbf{0.02}$}  & 4.26              & 2.39              & 1.49            & \colorbox{black!15}{$\mathbf{0.55}$} \\ \hline
			\multirow{3}{*}{\begin{tabular}[c]{@{}c@{}}New CI-DNN\\  $\Delta$SNR target: +5dB\\ for each stage\end{tabular}} & \stz{1 stage}    & 2.18              & \colorbox{black!15}{$\mathbf{0.02}$}  & 4.88              & 2.51              & 1.41            & \colorbox{black!15}{$\mathbf{0.54}$} \\ \cline{2-8} 
			& \stz{2 stage}    & 4.40              & \colorbox{black!15}{$\mathbf{0.03}$}  & 4.51              & 2.58              & 1.48            & \colorbox{black!15}{$\mathbf{0.55}$} \\ \cline{2-8} 
			& \stz{3 stage}    & 6.26              & 0.05             & 4.02              & 2.58              & 1.54            & \colorbox{black!15}{$\mathbf{0.54}$} \\ \hline
		\end{tabular}
	}
	\vspace*{-4mm}
	\caption{Performance for PED noise at {\bf $\text{SNR}\mathbf{=-5}\,\text{dB}$}; $\Delta$SNR and SSDR are measured in dB. {\bf Two best are \colorbox{black!15}{greyshaded}}.}
	\label{PED-noise-5db}
	\vspace*{-2mm}
\end{table}
\begin{table}[t!]
	\Large
	\centering
	\resizebox{1\linewidth}{!}{
		\begin{tabular}{|c|c||c|c||c|c||c|c|}
			\hline
			\multicolumn{2}{|c||}{\multirow{2}{*}{Method}}                                                                           & \multicolumn{2}{c||}{Noise Component} & \multicolumn{2}{c||}{Speech Component} & \multicolumn{2}{c|}{Total}        \\ \cline{3-8} 
			\multicolumn{2}{|c||}{}                                                                                                  & \stz{$\Delta$SNR}      & WLAKR             & SSDR              & PESQ$(\tilde{s})$  & PESQ$(\hat{s})$  & STOI            \\ \Xhline{3\arrayrulewidth}
			\multicolumn{2}{|c||}{\stz{LSA}}                                                                                               & 2.84             & 0.59              & \colorbox{black!15}{$\mathbf{5.55}$}   & 2.60              & 1.43            & 0.42            \\ \hline
			\multicolumn{2}{|c||}{\stz{SG}}                                                                                                & 3.17             & 0.60              & \colorbox{black!15}{$\mathbf{5.08}$}   & \colorbox{black!15}{$\mathbf{2.75}$}   & 1.40            & 0.41            \\ \hline
			\multicolumn{2}{|c||}{\stz{WF}}                                                                                                & 3.14             & 0.65              & 4.19              & \colorbox{black!15}{$\mathbf{2.99}$}   & 1.41            & 0.41            \\ \hline
			\multirow{3}{*}{\begin{tabular}[c]{@{}c@{}}Single DNN\\ clean target\end{tabular}}                                & \stz{1 stage}    & 5.87             & 0.13              & 3.68              & 2.20              & 1.59            & 0.53                \\ \cline{2-8} 
			& \stz{``2 stage"} & \colorbox{black!15}{$\mathbf{6.28}$}  & 0.20              & 3.75              & 2.15              & \colorbox{black!15}{$\mathbf{1.58}$} & 0.54            \\ \cline{2-8} 
			& \stz{``3 stage"} & \colorbox{black!15}{$\mathbf{6.50}$}  & 0.24              & 3.75              & 2.15              & \colorbox{black!15}{$\mathbf{1.59}$} & 0.54            \\ \hline
			\multirow{3}{*}{\begin{tabular}[c]{@{}c@{}}Single DNN\\ noisy target\end{tabular}}                                & \stz{1 stage}    & 1.99             & \colorbox{black!15}{$\mathbf{0.02}$}   & 4.38              & 2.45              & 1.43            & \colorbox{black!15}{$\mathbf{0.55}$} \\ \cline{2-8} 
			& \stz{``2 stage"} & 4.11             & \colorbox{black!15}{$\mathbf{0.02}$}   & 4.18              & 2.40              & 1.47            & \colorbox{black!15}{$\mathbf{0.56}$} \\ \cline{2-8} 
			& \stz{``3 stage"} & 5.76             & \colorbox{black!15}{$\mathbf{0.01}$}  & 4.02              & 2.33              & 1.52            & \colorbox{black!15}{$\mathbf{0.56}$} \\ \hline
			\multirow{3}{*}{\begin{tabular}[c]{@{}c@{}}New CI-DNN\\  $\Delta$SNR target: +5dB\\ for each stage\end{tabular}} & \stz{1 stage}    & 1.99             & \colorbox{black!15}{$\mathbf{0.02}$}   & 4.38              & 2.45              & 1.43            & \colorbox{black!15}{$\mathbf{0.55}$} \\ \cline{2-8} 
			& \stz{2 stage}    & 3.76             & 0.04              & 4.11              & 2.50              & 1.50            & \colorbox{black!15}{$\mathbf{0.55}$} \\ \cline{2-8} 
			& \stz{3 stage}    & 5.35             & 0.04              & 3.73              & 2.51              & 1.57            & \colorbox{black!15}{$\mathbf{0.55}$} \\ \hline
		\end{tabular}
	}
	\vspace*{-4mm}
	\caption{Performance for CAFE noise at {\bf $\text{SNR}\mathbf{=-5}\,\text{dB}$}; $\Delta$SNR and SSDR are measured in dB. {\bf Two best are \colorbox{black!15}{greyshaded}}.}
	\label{CAFE-noise-5db}
	\vspace*{-1.8mm}
\end{table}
\begin{table}[t!]
	\Large
	\centering
	\resizebox{1\linewidth}{!}{
		\begin{tabular}{|c|c||c|c||c|c||c|c|}
			\hline
			\multicolumn{2}{|c||}{\multirow{2}{*}{Method}}                                                                           & \multicolumn{2}{c||}{Noise Component} & \multicolumn{2}{c||}{Speech Component} & \multicolumn{2}{c|}{Total}        \\ \cline{3-8} 
			\multicolumn{2}{|c||}{}                                                                                                 & \stz{$\Delta$SNR}        & WLAKR            & SSDR              & PESQ$(\tilde{s})$  & PESQ$(\hat{s})$  & STOI            \\ \Xhline{3\arrayrulewidth}
			\multicolumn{2}{|c||}{\stz{LSA}}                                                                                               & 5.10              & 0.59             & \colorbox{black!15}{$\mathbf{8.20}$}   & 3.03              & \colorbox{black!15}{$\mathbf{1.68}$} & 0.56            \\ \hline
			\multicolumn{2}{|c||}{\stz{SG}}                                                                                                & 5.09              & 0.60             & \colorbox{black!15}{$\mathbf{7.75}$}   & \colorbox{black!15}{$\mathbf{3.22}$}   & 1.64            & 0.56            \\ \hline
			\multicolumn{2}{|c||}{\stz{WF}}                                                                                                & \colorbox{black!15}{$\mathbf{6.42}$}   & 0.69             & 6.63              & \colorbox{black!15}{$\mathbf{3.37}$}   & \colorbox{black!15}{$\mathbf{1.69}$} & 0.55            \\ \hline
			\multirow{3}{*}{\begin{tabular}[c]{@{}c@{}}Single DNN\\ clean target\end{tabular}}                                & \stz{1 stage}    & 6.07              & \colorbox{black!15}{$\mathbf{0.01}$}  & 7.09              & 2.55              & 1.64            & \colorbox{black!15}{$\mathbf{0.65}$} \\ \cline{2-8} 
			& \stz{``2 stage"} & 4.47              & 0.15             & 7.11              & 2.43              & 1.57            & 0.64            \\ \cline{2-8} 
			& \stz{``3 stage"} & 4.73              & 0.11             & 7.19              & 2.43              & 1.57            & 0.64            \\ \hline
			\multirow{3}{*}{\begin{tabular}[c]{@{}c@{}}Single DNN\\ noisy target\end{tabular}}                                & \stz{1 stage}    & 2.14              & \colorbox{black!15}{$\mathbf{0.01}$}  & 7.42              & 2.59              & 1.40            & \colorbox{black!15}{$\mathbf{0.65}$} \\ \cline{2-8} 
			& \stz{``2 stage"} & 3.48              & \colorbox{black!15}{$\mathbf{0.03}$}  & 7.36              & 2.60              & 1.48            & \colorbox{black!15}{$\mathbf{0.65}$} \\ \cline{2-8} 
			& \stz{``3 stage"} & 4.57              & 0.07             & 7.30              & 2.56              & 1.55            & \colorbox{black!15}{$\mathbf{0.65}$} \\ \hline
			\multirow{3}{*}{\begin{tabular}[c]{@{}c@{}}New CI-DNN\\  $\Delta$SNR target: +5dB\\ for each stage\end{tabular}} & \stz{1 stage}    & 2.14              & \colorbox{black!15}{$\mathbf{0.01}$}  & 7.42              & 2.59              & 1.40            & \colorbox{black!15}{$\mathbf{0.65}$} \\ \cline{2-8} 
			& \stz{2 stage}    & 4.46              & \colorbox{black!15}{$\mathbf{0.03}$}  & 7.34              & 2.72              & 1.51            & \colorbox{black!15}{$\mathbf{0.66}$} \\ \cline{2-8} 
			& \stz{3 stage}    & \colorbox{black!15}{$\mathbf{6.50}$}   & 0.04             & 7.00              & 2.75              & 1.66            & \colorbox{black!15}{$\mathbf{0.65}$} \\ \hline
		\end{tabular}
	}
	\vspace*{-4mm}
	\caption{Performance for {\bf unseen} BUS noise at {\bf $\text{SNR}\mathbf{=-5}\,\text{dB}$}; $\Delta$SNR and SSDR are measured in dB. {\bf Two best are \colorbox{black!15}{greyshaded}}.}
	\label{BUS-noise-5db}
	\vspace*{-6mm}
\end{table}
In this paper, we proposed serially concatenated identical DNNs (CI-DNNs), where each basic DNN module (stage) can offer some moderate enhancement of the input. Our proposed CI-DNNs outperform the classical spectral weighting rules both in total speech quality and speech intelligibility. The CI-DNN also shows more balanced performance than the conventional clean-target single DNN. Comparing with the noisy-target single DNN, our proposed CI-DNN offers quite similar or even a bit better performance concerning total PESQ$(\hat{s})$, but with only $1/2$ or $1/3$ of the trainable weights. Under a comparable noise and speech {\it component} quality, our proposed CI-DNNs also generalize better to an unseen noise type by offering higher total PESQ$(\hat{s})$ and SNR improvement.



\bibliographystyle{IEEEbib}
\ninept
\vspace*{-4mm}
\bibliography{main}

\begin{thebibliography}{10}

\bibitem{Ephraim1984}
{Y}. {E}phraim and {D}. {M}alah,
\newblock ``{S}peech {E}nhancement {U}sing a {M}inimum {M}ean-{S}quare {E}rror
  {S}hort-{T}ime {S}pectral {A}mplitude {E}stimator,''
\newblock {\em IEEE T-ASSP}, vol. 32, no. 6, pp. 1109--1121, Dec. 1984.

\bibitem{Ephraim1985}
{Y}. {E}phraim and {D}. {M}alah,
\newblock ``{S}peech {E}nhancement {U}sing a {M}inimum {M}ean-{S}quare {E}rror
  {L}og-{S}pectral {A}mplitude {E}stimator,''
\newblock {\em IEEE T-ASSP}, vol. 33, no. 2, pp. 443--445, Apr. 1985.

\bibitem{Scalart1996}
{P}. {S}calart and {J}.~{V}. {F}ilho,
\newblock ``{S}peech {E}nhancement {B}ased on {A} {P}riori {S}ignal to {N}oise
  {E}stimation,''
\newblock in {\em {P}roc. of {ICASSP}}, Atlanta, GA, USA, May 1996, pp.
  629--632.

\bibitem{Lotter2005}
{T}. {L}otter and {P}. {V}ary,
\newblock ``{S}peech {E}nhancement by {MAP} {S}pectral {A}mplitude {E}stimation
  {U}sing a {S}uper-{G}aussian {S}peech {M}odel,''
\newblock {\em EURASIP Journal on Applied Signal Processing}, vol. 2005, no. 7,
  pp. 1110--1126, 2005.

\bibitem{Cohen2005a}
{I}. {C}ohen,
\newblock ``{S}peech {E}nhancement {U}sing {S}uper-{Gauss}ian {S}peech {M}odels
  and {N}oncausal {A} {P}riori {SNR} {E}stimation,''
\newblock {\em Speech Commun.}, vol. 47, no. 3, pp. 336--350, Nov. 2005.

\bibitem{Gerkmann2008b}
{T}. {G}erkmann, {C}. {B}reithaupt, and {R}. {M}artin,
\newblock ``{I}mproved {A} {P}osteriori {S}peech {P}resence {P}robability
  {E}stimation {B}ased on a {L}ikelihood {R}atio with {F}ixed {P}riors,''
\newblock {\em IEEE T-ASLP}, vol. 16, no. 5, pp. 910--919, July 2008.

\bibitem{Suhadi2011}
{S}. {S}uhadi, {C}. {L}ast, and {T}. {F}ingscheidt,
\newblock ``{A} {D}ata-{D}riven {A}pproach to {A} {P}riori {SNR}
  {E}stimation,''
\newblock {\em IEEE T-ASLP}, vol. 19, no. 1, pp. 186--195, Jan. 2011.

\bibitem{Elshamy2015}
{S}. {E}lshamy, {N}. {M}adhu, {W}.~{J}. {T}irry, and {T}. {F}ingscheidt,
\newblock ``{A}n {I}terative {S}peech {M}odel-{B}ased {A} {P}riori {SNR}
  {E}stimator,''
\newblock in {\em {P}roc. of {I}nterspeech}, Dresden, Germany, Sept. 2015, pp.
  1740--1744.

\bibitem{malah1999tracking}
{D}. {M}alah, {R}.~{V}. {C}ox, and {A}.~{J}. {A}ccardi,
\newblock ``{T}racking {S}peech-presence {U}ncertainty to {I}mprove {S}peech
  {E}nhancement in {N}on-stationary {N}oise {E}nvironments,''
\newblock in {\em {P}roc. of {ICASSP}}, Phoenix, AZ, USA, Mar. 1999, pp.
  789--792.

\bibitem{du2016regression}
{J}. {D}u, {Y}. {T}u, {L}.~{R}. Dai, and {C}.~{H}. {L}ee,
\newblock ``{A} {R}egression {A}pproach to {S}ingle-{C}hannel {S}peech
  {S}eparation via {H}igh-{R}esolution {D}eep {N}eural {N}etworks,''
\newblock {\em IEEE/ACM T-ASLP}, vol. 24, no. 8, pp. 1424--1437, Apr. 2016.

\bibitem{wang2014training}
{Y}. {W}ang, {A}. {N}arayanan, and {D}.~{L}. {W}ang,
\newblock ``{O}n {T}raining {T}argets for {S}upervised {S}peech {S}eparation,''
\newblock {\em IEEE/ACM T-ASLP}, vol. 22, no. 12, pp. 1849--1858, Dec. 2014.

\bibitem{wang2015deep}
{Y}. {W}ang and {D}.~{L}. {W}ang,
\newblock ``{A} {D}eep {N}eural {N}etwork for {T}ime-{D}omain {S}ignal
  {R}econstruction,''
\newblock in {\em {P}roc. of {ICASSP}}, Brisbane, QLD, Australia, Aug. 2015,
  pp. 4390--4394.

\bibitem{erdogan2015phase}
{H}. {E}rdogan, {J}.~{R}. {H}ershey, {S}. {W}atanabe, and {J}. {L}e {R}oux,
\newblock ``{P}hase-{S}ensitive and {R}ecognition-{B}oosted {S}peech
  {S}eparation {U}sing {D}eep {R}ecurrent {N}eural {N}etworks,''
\newblock in {\em {P}roc. of {ICASSP}}, Brisbane, QLD, Australia, Aug. 2015,
  pp. 708--712.

\bibitem{weninger2014discriminatively}
{F}. {W}eninger, {J}.~{R}. {H}ershey, {J}. {L}e {R}oux, and {B}. {S}chuller,
\newblock ``{D}iscriminatively {T}rained {R}ecurrent {N}eural {N}etworks for
  {S}ingle-{C}hannel {S}peech {S}eparation,''
\newblock in {\em {P}roc. of 2nd IEEE {GlobalSIP}}, Atlanta, GA, USA, May 2014,
  pp. 577--581.

\bibitem{williamson2016complex}
{D}.~{S}. {W}illiamson, {Y}. {W}ang, and {D}.~{L}. {W}ang,
\newblock ``{C}omplex {R}atio {M}asking for {M}onaural {S}peech {S}eparation,''
\newblock {\em IEEE/ACM T-ASLP}, vol. 24, no. 3, pp. 483--492, Mar. 2016.

\bibitem{xu2014experimental}
{Y}. {Xu}, {J}. {D}u, {L}.~{R}. {D}ai, and {C}.~{H}. {L}ee,
\newblock ``{A}n {E}xperimental {S}tudy on {S}peech {E}nhancement {B}ased on
  {D}eep {N}eural {N}etworks,''
\newblock {\em IEEE Signal Processing Letters}, vol. 21, no. 1, pp. 65--68,
  Nov. 2014.

\bibitem{wang2013towards}
{Y}. {W}ang and {D}.~{L}. {W}ang,
\newblock ``{T}owards {S}caling up {C}lassification-{B}ased {S}peech
  {S}eparation,''
\newblock {\em IEEE T-ASLP}, vol. 21, no. 7, pp. 1381--1390, Mar. 2013.

\bibitem{xu2015regression}
{Y}. {X}u, {J}. {D}u, {L}.~{R}. {D}ai, and {C}.~{H}. {L}ee,
\newblock ``{A} {R}egression {A}pproach to {S}peech {E}nhancement {B}ased on
  {D}eep {N}eural {N}etworks,''
\newblock {\em IEEE/ACM T-ASLP}, vol. 23, no. 1, pp. 7--19, Jan. 2015.

\bibitem{elshamy2018dnn}
{S}. {E}lshamy, {N}. {M}adhu, {W}. {T}irry, and {T}. {F}ingscheidt,
\newblock ``{DNN}-{S}upported {S}peech {E}nhancement {W}ith {C}epstral
  {E}stimation of {B}oth {E}xcitation and {E}nvelope,''
\newblock {\em IEEE/ACM T-ASLP}, vol. 26, no. 12, pp. 2460--2474, 2018.

\bibitem{gao2016snr}
{T}. {G}ao, {J}. {D}u, {L}.~{R}. {D}ai, and {C}.~{H}. {L}ee,
\newblock ``{SNR}-{B}ased {P}rogressive {L}earning of {D}eep {N}eural {N}etwork
  for {S}peech {E}nhancement.,''
\newblock in {\em {P}roc. of {I}nterspeech}, San Francisco, CA, USA, Sep. 2016,
  pp. 3713--3717.

\bibitem{Shu2018iwaenc}
{X}. {S}hu, {Y}. {Z}hou, and {Y}. {C}ao,
\newblock ``{A} {N}ew {S}peech {E}nhancement {A}pproach {B}ased on
  {P}rogressive {D}eep {N}eural {N}etworks,''
\newblock in {\em {P}roc. of {IWAENC}}, Tokyo, Japan, Sep. 2018, pp. 191--195.

\bibitem{ITU56}
{I}{T}{U},
\newblock {\em {O}bjective {M}easurement of {A}ctive {S}peech {L}evel},
\newblock {International Telecommunication Standardization Sector (ITU-T),\,
  Rec. P.56}, Dec. 2011.

\bibitem{veit2016residual}
{A}. {V}eit, {M}.~{J}. {W}ilber, and {S}. {B}elongie,
\newblock ``{R}esidual {N}etworks {B}ehave {L}ike {E}nsembles of {R}elatively
  {S}hallow {N}etworks,''
\newblock in {\em {P}roc. of {NIPS}}, Barcelona, Spain, Dec. 2016, pp.
  550--558.

\bibitem{grid_corpus}
{M}. {C}ooke, {J}. {B}arker, {S}. {C}unningham, and {X}. {S}hao,
\newblock ``{A}n {A}udio-{V}isual {C}orpus for {S}peech {P}erception and
  {A}utomatic {S}peech {R}ecognition,''
\newblock {\em The Journal of the Acoustical Society of America}, vol. 120, no.
  5, pp. 2421--2424, Jun. 2006.

\bibitem{chime3}
{J}. {B}arker, {R}. {M}arxer, {E}. {V}incent, and {S}. Watanabe,
\newblock ``The {T}hird {`C}{H}i{M}{E'} {S}peech {S}eparation and {R}ecognition
  {C}hallenge: {D}ataset, {T}ask and {B}aselines,''
\newblock in {\em {P}roc. of {ASRU}}, Scottsdale, AZ, USA, Feb. 2015, pp.
  504--511.

\bibitem{samy_SNR}
{S}. Elshamy, {N}. Madhu, {W}. Tirry, and {T}. Fingscheidt,
\newblock ``{I}nstantaneous {A} {P}riori {SNR} {E}stimation by {C}epstral
  {E}xcitation {M}anipulation,''
\newblock {\em IEEE/ACM T-ASLP}, vol. 25, no. 8, pp. 1592--1605, Aug. 2017.

\bibitem{ITU862}
{I}{T}{U},
\newblock {\em {Perceptual Evaluation of Speech Quality (PESQ): An Objective
  Method for End-To-End Speech Quality Assessment of Narrow-Band Telephone
  Networks and Speech Codecs}},
\newblock {International Telecommunication Standardization Sector (ITU-T), Rec.
  P.862}, Feb. 2001.

\bibitem{ITU1110}
{I}{T}{U},
\newblock {\em {W}ideband {H}ands-{F}ree {C}ommunication in {M}otor
  {V}ehicles},
\newblock {International Telecommunication Standardization Sector (ITU-T), Rec.
  P.1110}, Jan. 2015.

\bibitem{ITU1130}
{I}{T}{U},
\newblock {\em {Subsystem Requirements for Automotive Speech Services}},
\newblock {International Telecommunication Standardization Sector (ITU-T), Rec.
  P.1130}, Jun. 2015.

\bibitem{yuDSP2011a}
{H}. {Y}u and {T}. {F}ingscheidt,
\newblock ``{A} {F}igure of {M}erit for {I}nstrumental {O}ptimization of
  {N}oise {R}eduction {A}lgorithms,''
\newblock in {\em {P}roc. of 5th Biennial Workshop on DSP for In-Vehicle
  Systems}, Kiel, Germany, Sep. 2011, pp. 1--8.

\bibitem{taal2010short}
{C}.~{H}. Taal, {R}.~{C}. {H}endriks, {R}. {H}eusdens, and {J}. {J}ensen,
\newblock ``{A} {S}hort-{T}ime {O}bjective {I}ntelligibility {M}easure for
  {T}ime-{F}requency {W}eighted {N}oisy {S}peech,''
\newblock in {\em {P}roc. of {ICASSP}}, Dallas, TX, USA, Jun. 2010, pp.
  4214--4217.

\bibitem{Martin2001}
{R}. {M}artin,
\newblock ``{N}oise {P}ower {S}pectral {D}ensity {E}stimation {B}ased on
  {O}ptimal {S}moothing and {M}inimum {S}tatistics,''
\newblock {\em IEEE T-SAP}, vol. 9, no. 5, pp. 504--512, Jul. 2001.

\end{thebibliography}
\end{document}